\documentclass[runningheads]{llncs}
\usepackage[T1]{fontenc}
\usepackage{graphicx}
\usepackage{verbatim}
\usepackage[dvipsnames]{xcolor}
\usepackage{wrapfig}
\PassOptionsToPackage{hyphens}{url}
\usepackage{hyperref}
\usepackage{float}
\usepackage{caption}
\usepackage{subcaption}

\usepackage{listings}
\begin{document}
%
\title{Speeding up SQL subqueries via decoupling of non-correlated predicate (extended version)}

%
%
\author{
Dmitrii Radivonchik\inst{1}\orcidID{0009-0005-9885-6257} 
\and
Yakov Kuzin\inst{2}\orcidID{0009-0009-2961-6021} \and
Anton Chizhov \inst{2}\orcidID{0009-0007-3042-345X} 
\and
Dmitriy 
Shcheka\inst{2}\orcidID{0009-0000-1519-8740} \and
Mikhail Firsov\inst{2}\orcidID{0000-0001-6739-7303} \and
Kirill Smirnov\inst{2}\orcidID{0000-0003-4727-3455} \and
George Chernishev\inst{2}\orcidID{0000-0002-4265-9642}}
\authorrunning{D. Radivonchik et al.}
%

\institute{
Independent, Amsterdam, Netherlands\\
\email{dmitry.radivonchik@gmail.com}
\and
Saint-Petersburg University, Russia\\
\email{\{yakov.s.kuzin, anton.i.chizhov, dmitriy.v.shcheka, mikhail.a.firsov, kirill.k.smirnov, chernishev\}@gmail.com}
}
\maketitle              
\begin{abstract}

In this paper, we discuss a novel technique for processing correlated subqueries in SQL. The core idea is to isolate the non-correlated part of the predicate and use it to reduce the number of evaluations of the correlated part. We begin by providing an overview of several classes of queries that may benefit from this technique. For each class, we propose a potential rewrite and discuss the conditions under which it is advantageous. Next, we address the evaluation aspects of the proposed rewrites: 1) we describe our approach to adapting the block-based Volcano query processing model, and 2) we discuss the benefits of implementing that technique within a position-enabled column-store with late materialization support. Finally, we present a simple cost model that allows estimation of the benefits of said rewrites.

Our evaluation has a quantitative part and a qualitative part. The former focuses on studying the impact of non‑correlated predicate selectivity on our technique. The latter identifies the limitations of our approach by comparing it with alternative approaches available in existing systems. Overall, experiments conducted using PosDB (a position‑enabled column-store) and PostgreSQL demonstrated that, under suitable conditions, our technique can achieve a 5$\times$ improvement.

\keywords{Query Processing \and Column-stores \and Subquery Processing \and Late Materialization \and PosDB.}
\end{abstract}
%
%
%
\section{Introduction}

In SQL, a subquery is a query that resides inside some other query. The former one is referred to as inner, and the latter one is called outer. Subqueries are an important part of data analysis~--- for example, in the TPC-H benchmark~\cite{TPCH}, 14 out of the 22 presented queries have subqueries in them.

Evaluating subqueries can be very costly, especially when dealing with so-called \textit{correlated} subqueries. Such subqueries require complete reevaluation of the inner query for each row of the outer one. Therefore, optimizing subqueries is a relevant task, and a successful solution can bring significant benefits to the industry.

In this paper we propose a novel technique for processing correlated subqueries in SQL. We consider a special case when the \texttt{WHERE} clause contains a compound predicate, which consists of correlated and non-correlated parts. In some cases it is possible to isolate the non-correlated part of the predicate and use it to reduce the number of evaluations of the correlated part.

To achieve that, we identify several classes of correlated subqueries, present possible rewrites for them and discuss the conditions under which they are advantageous. We focus on a particular class featuring an \texttt{OR} predicate and present an approach to adapting the block-based Volcano query processing model for it.

The approach is discussed in the context of a position-enabled column-store with late materialization support. Unlike regular column-stores, position-enabled systems feature fluent position handling, enabling the query engine to explicitly use positions during query execution. This allows for a number of optimizations and techniques that offer various benefits for query processing. 

Late materialization~\cite{columns_tutorial,Abadi:2013:DIM:2602024} is the idea that fuels a large number of such techniques. The idea is to operate on positions and delay tuple reconstruction for as long as possible. Such an approach allows query engine to significantly conserve I/O bandwidth and to reduce the CPU processing load for appropriate queries.

Although the proposed technique is applicable and can be useful for row stores as well, we focus specifically on the position‑enabled column‑store case. The power of such systems stems from the ability to compose long chains of positional operators. The longer the results are processed as positions, the greater the chance the query executor can take advantage of the reduced number of rows in the final output, yielding an overall cheaper query plan.

Thus, this work introduces, for the first time, a positional operator that enables constructing positional operator chains for subquery case.

We also present a simple cost model that allows for estimating the benefits of our rewrites.

Finally, we perform an evaluation of the proposed technique, which consisted of quantitative and qualitative parts. The quantitative part shows the impact of non‑correlated predicate selectivity on our technique. It identifies the cases (for the chosen class of queries~--- a subquery with the \texttt{OR} predicate) where the technique is beneficial. Together with the cost models mentioned above, this study lays the groundwork for an optimizer that can decide whether to apply this technique for a given query.

The qualitative part aims to position our technique among other approaches and further delineate its applicability. To that end, we analyze alternative query plans produced by an industrial DBMS. Subquery optimization is a well‑studied area with many competing techniques that are often high‑performing; therefore, we discuss cases where our technique may underperform in a broader sense.


Overall, the contribution of this paper is the following:

\begin{enumerate}
    \item A classification of correlated subquery types that have \texttt{WHERE} clause consisting of the correlated and non-correlated parts. This classification is supplied with equivalent rewrites for each class.
    
    \item A technique for evaluating the rewrite of one of such classes. The idea is to isolate a non-correlated part of the predicate and use it to reduce the number of evaluations of the correlated part.
    
    \item An approach to adapting the block-based Volcano query processing model for this technique. We consider this in the context of a position-enabled column-store. As a result, we have proposed a novel positional operator that enables constructing positional operator chains for the subquery case.
    
    \item A simple cost model that allows for estimating the benefits of rewrites. It can serve as the groundwork for an optimizer that can decide whether to apply our technique for a given query.
    
    \item An evaluation: a comparison with PostgreSQL which demonstrated that our technique can achieve a 5$\times$ improvement and a discussion of applicability limits of our technique.
\end{enumerate}

\section{Related Work}

The optimization of SQL subqueries has been a long-standing challenge in da\-ta\-ba\-se systems, leading to a rich body of work on both logical transformations and execution strategies. Our proposed technique builds upon this foundation, particularly drawing inspiration from research on decorrelation and efficient evaluation of correlated queries. The following section surveys key related work that provides a base for our approach.

Elhemali et al.~\cite{10.1145/1247480.1247598} provide an overview of the subquery execution strategies implemented in Microsoft SQL Server, categorizing them into navigational (e.g., forward and reverse lookup) and set-oriented approaches. Their work details the process of decorrelation to enable efficient join-based execution but also acknowledges scenarios where decorrelation is infeasible or undesirable, advocating for a cost-based choice among multiple strategies. They further discuss optimizations for non-decorrelated execution, such as caching, prefetching, and batch sorting for nested iterations. While their work offers a broad industrial perspective on handling various subquery forms, our technique specifically targets a narrower class of queries featuring a compound predicate (NC OR C) within a correlated subquery. We propose a novel rewrite that isolates the non-correlated (NC) part to act as a pre-filter, a strategy complementary to their navigational optimizations but focused on reducing the computational overhead of the correlated (C) part within a disjunction, an area their work identifies as particularly challenging for decorrelation.

Bellamkonda et al.~\cite{10.14778/1687553.1687563} present a suite of subquery optimizations in Oracle, focusing on techniques like subquery coalescing and subquery removal using window functions. Their approach is primarily based on query transformation, rewriting complex nested queries into more efficient, set-oriented forms. For instance, they coalesce multiple subqueries with overlapping logic into a single one and replace correlated aggregates with window functions that compute results over partitions, thereby avoiding repeated evaluations. While these transformations are powerful and can lead to significant performance gains, they are applied during the logical optimization phase and require a complex cost-based transformation framework to navigate the search space of equivalent query forms. Our work differs by focusing on a specific class of queries with compound correlated predicates. Our execution-level strategy is complementary to Oracle's logical transformations and could be integrated to handle cases where full decorrelation is not possible or beneficial.

Abeysinghe, He and Rompf~\cite{10.1145/3514221.3517889} address the challenge of incrementally evaluating nested aggregate queries with correlations, a problem that becomes critical in low-latency streaming contexts. They propose Relative Partial Aggregate Indexes (RPAI), a tree-based data structure that enables efficient range shifts and aggregate queries in logarithmic time. This approach significantly outperforms state-of-the-art systems like DBToaster, which often fall back to full recomputation for such queries. While their work focuses on incremental view maintenance and streaming scenarios, it shares a common goal with ours: reducing the computational overhead of correlated subqueries. However, their technique is specialized for aggregate-heavy workloads and recursive delta processing, whereas our approach targets a broader class of predicate-based subqueries in traditional column-stores and uses late materialization and predicate isolation to minimize redundant computation.

Zhao et al.~\cite{10.1145/3589330} tackle the problem of query re-optimization, identifying a key weakness in existing approaches: their heavy reliance on an initial, often suboptimal, global physical plan for dividing the query into re-optimization units. They propose QuerySplit, a proactive algorithm that bypasses the initial plan entirely. Instead, it extracts subqueries directly from the logical plan based on primary-foreign-key relationships, forming re-optimization units that are less likely to be misled by cardinality estimation errors. A cost function then prioritizes the execution of subqueries with low estimated cost and small output size, thereby delaying costly large joins and increasing the chance they will operate on reduced inputs. While their work focuses on adaptive join ordering and physical plan generation in the face of estimation errors, it shares a common theme with our technique: the strategic decomposition of a query into more manageable and efficiently evaluable parts. However, QuerySplit operates at the level of entire join subqueries for re-optimization, whereas our approach focuses on a finer-grained, predicate-level isolation within correlated subqueries to minimize redundant computation during their evaluation.

Kim and Madden~\cite{10.1145/3654961} introduce an original tagged execution model to optimize queries with complex disjunctive predicates. Their approach groups tuples into subrelations based on which predicates they satisfy, tagging them with this semantic information. These tags enable operators to avoid redundant work and achieve predicate pushdown for disjunctions, a known challenge for traditional optimizers. While their technique is powerful for general disjunctive queries, it introduces overhead from tag management and requires significant changes to the query engine's core operators. In contrast, our proposed technique targets a more specific, but common, case: correlated subqueries with a compound (AND/OR) predicate. Our approach requires less invasive changes to the execution model, making it a more lightweight optimization that is complementary to general frameworks like tagged execution, especially within the context of a position-enabled column-store with late materialization.

Bruno et al.~\cite{10.1145/3722212.3724448} present a universal framework for query decorrelation within the Microsoft Fabric Data Warehouse, building upon and extending the foundational work of Galindo-Legaria and Joshi~\cite{10.1145/1247480.1247598}. Their approach formalizes the problem using the APPLY operator and provides a set of algebraic equivalences to remove correlation. A key contribution is their system's ability to avoid unnecessary expression duplication through heuristic ``correlation pulling'' and specialized equivalences, which is crucial for maintaining plan efficiency. While their technique is powerful for general decorrelation, transforming a correlated plan into a fully set-oriented one, it operates at the logical algebra level and requires significant integration into the query optimizer. Once again, in contrast, our technique is more surgical, targeting a specific but common pattern, making it a valid complementary approach.


\section{The Proposed Technique}


\subsection{Sketch of the Technique}\label{sec:sketch}

Let's illustrate the idea of the approach with a concrete example. Consider the query presented in Fig.~\ref{listing:sq1}. It searches for students who work as teaching assistants and either work in department \texttt{Dep1} or earn more than some other employee.

\begin{lstlisting}[label={listing:sq1}, caption={An example subquery}]
SELECT S.name, S.salary
    FROM students AS S
    WHERE S.name IN (
        SELECT E.name
            FROM employees AS E
            WHERE E.department = 'Dep1' OR
                  E.salary < S.salary
    );
\end{lstlisting}

The most straightforward approach to executing this query is as follows. For each row of the \texttt{students} table, iterate over the rows of the \texttt{employees} while checking the compound predicate. There are other advanced ways to execute such subqueries, they will be discussed later in the evaluation section.

Our idea is to isolate the non-correlated part and use it as a filter that allows ``skipping'' some iterations over the inner table. To achieve this, the query should be rewritten as presented in Fig.~\ref{listing:sq1-rewritten}.

\begin{lstlisting}[label={listing:sq1-rewritten}, caption={Query after rewrite}]
SELECT S.name, S.salary
    FROM students AS S
    WHERE S.name IN (
        SELECT E.name
            FROM employees AS E
            WHERE E.department = 'Dep1'
    ) OR S.name IN (
        SELECT E.name
            FROM employees AS E
            WHERE E.salary < S.salary
    );
\end{lstlisting}

This query contains two subqueries in the \texttt{WHERE} clause. The first one is a non-correlated subquery, whose result does not change with each row of the \texttt{students} table. Therefore, it can be computed only once and using a single pass over the \texttt{employees} table.

As a result, it allows the following evaluation technique: for each record from the \texttt{students} table, first check the ``cheap'' non-correlated subquery, which is cached. Then, if the result is false, run costly correlated subquery, which will make a full pass over the \texttt{students} table.

In the next sections, we will list and systematize query classes which can be processed by this technique, as well as consider implementation details.

\subsection{Query Classes}\label{sec:classes}

To assess the scope of applicability, we analyzed possible queries. In these queries we varied: 1) the logical operator connecting two predicates in the inner query, and 2) the subquery predicate. For the former we decided to consider \texttt{OR} and \texttt{AND} predicates, while for the latter we restricted ourselves to \texttt{IN} and \texttt{<SOME}. We select these predicates because they demonstrate rewrites that generalize to other cases and will help highlight important implementation details later. Overall, the following query classes are distinguished:

\begin{enumerate}
    \item \texttt{SELECT ... WHERE X IN (SELECT ... WHERE NC AND C)}
    \item \texttt{SELECT ... WHERE X IN (SELECT ... WHERE NC OR C)}
    \item \texttt{SELECT ... WHERE X <SOME (SELECT ... WHERE NC AND C)}
    \item \texttt{SELECT ... WHERE X <SOME (SELECT ... WHERE NC OR C)}
\end{enumerate}

To conserve space we denote a \textit{non-correlated} predicate as \texttt{NC} and \textit{correlated} as C. 

Classes 1 and 3 can be rewritten in the form of \texttt{OP NC AND OP (NC AND C)}, where \texttt{OP} is either \texttt{IN} or \texttt{<SOME}. Note that it was needed to put the \texttt{NC} predicate both inside the correlated and non-correlated parts of the query to ensure the correctness of the rewrite.

The evaluation of the rewritten version does not require any modifications to the operator code of a query engine. It is sufficient to reorder operators inside the query plan as follows:

\begin{enumerate}
    \item Pre-filter the inner table by eliminating rows that do not satisfy the \texttt{NC} predicate.
    \item For each row of the outer table check the \texttt{C} predicate over each row of the pre-filtered inner table.
\end{enumerate}

Pre-filtering can substantially reduce the size of the intermediate results. However, there is an overhead: the \texttt{NC} predicate will be checked twice. Therefore, there will be a trade-off between the selectivity of \texttt{NC} and its cost.

Classes 2 and 4 can be rewritten in the form of \texttt{OP NC OR OP C}, where \texttt{OP} is either \texttt{IN} or \texttt{<SOME} operator. The evaluation of the rewritten version is different. It was discussed earlier, in Section~\ref{sec:sketch}. To implement this case, a simple modification of a query plan will not suffice, one needs to perform rework in the query engine core.

\subsection{Implementation}

We are going to implement our technique inside position-enabled column-store PosDB\footnote{https://pos-db.com/}. First, we are going to introduce basics, closely following the reference~\cite{10.1007/978-3-031-49333-1_14} (see Section 3, ``Background'').

\begin{figure*}[t]
\centering
\begin{subfigure}{.29\textwidth}
    \centering
    \includegraphics[width=0.75\textwidth]{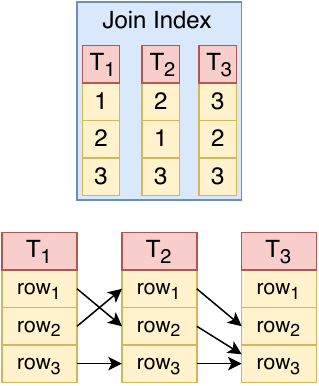}
    \caption{Example of join index}
    \label{fig:join-index}
\end{subfigure}
\begin{subfigure}{.7\textwidth}
    \centering
    \includegraphics[width=0.9\textwidth]{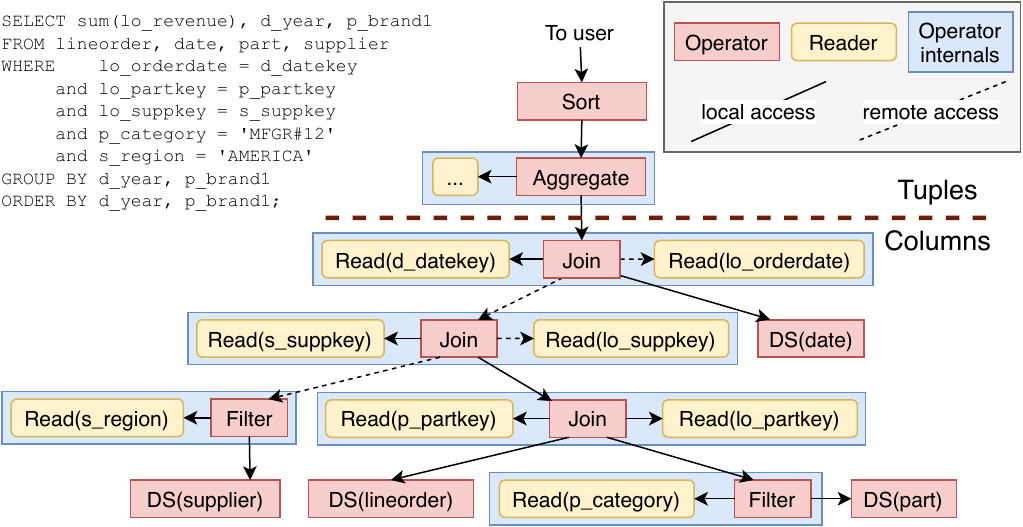}
    \caption{Query plan example}
    \label{fig:plan}
\end{subfigure}
\caption{PosDB internals}
\label{fig:test}
\end{figure*}

PosDB uses the block-based Volcano model~\cite{Graefe:1993:QET:152610.152611}. Its core feature is two types of intermediate representations: tuple- and position-based. In the tuple-based representation, operators pass blocks of value tuples. This type of representation is similar to most existing DBMSs. Conversely, position-based representation is a characteristic feature of PosDB. In the positional form, intermediates are represented by a generalized join index~\cite{Valduriez:1987:JI:22952.22955} which stores an array of record indices, i.e., positions, for each table it covers (top of Fig.~\ref{fig:join-index}). Tuples are encoded using rows in the join index.

Most operators in PosDB are either positional or tuple-based, with positional ones having specialized Reader entities for reading values of individual table attributes. The query plan in PosDB is divided into positional and tuple parts, and the moment of converting positions into tuples is called materialization. Materialization is to be performed at some moment of query plan, since the user needs tuples and not positions. It can be performed by either a special \texttt{Materialize} operator or by some operators, such as an aggregation operator. Finally, the \texttt{Datasource} (DS) operator initializes data flows with join index of unit cardinality~--- a list of positions belonging to each table. An example query plan is presented in Fig.~\ref{fig:plan}.

A comprehensive description of the system can be found in the recent papers~\cite{10.1145/3632410.3632422,DBLP:conf/dolap/ChernishevGGK022}.

Despite that there are no specialized operators in the relational algebra~\cite{DBLP:books/mg/SKS20} for expressing subqueries, implementing them inside query engine requires significant modifications. This is especially true regarding correlated subqueries, which need to track dependence between outer and inner parameters. This was the first modification of the Volcano model, which was needed to support correlated subquery processing. In further figures we denote multiple invocations of the inner plan by double arrows.

\begin{figure}[htp]
    \begin{center}
        \includegraphics[width=\linewidth]{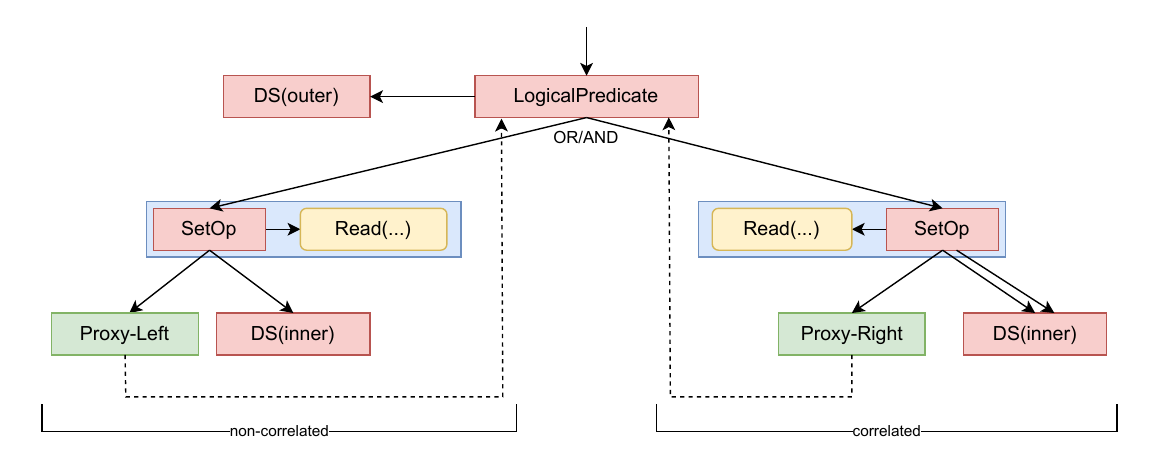}
    \end{center}
    \caption{LP architecture}    \label{fig:lp_architecture}
\end{figure}

Next, to implement our technique itself we propose the LP operator (Logical Predicate). The principal scheme of our approach is presented in Fig.~\ref{fig:lp_architecture}.

This operator coordinates the processing of the whole query by invoking different parts of the plan. Its state is comprised of: 1) the DS operator, which provides positional data for outer table, 2) a predicate, which constitutes the outer \texttt{WHERE}, 3) two child nodes, which represent subquery parts.

\begin{figure}[t]
    \begin{center}
        \includegraphics[width=0.6\linewidth]{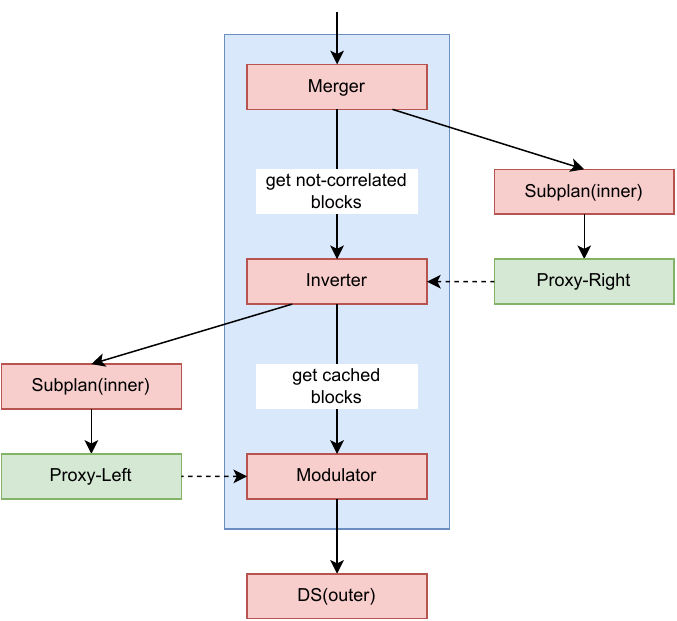}
    \end{center}
    \caption{LP data flow}
    \label{fig:lp_implementation}
\end{figure}

To support subquery processing in PosDB, we have implemented the positional \texttt{SetOp} operator. It expresses the logic of any set-based subquery predicate~--- \texttt{IN}, \texttt{<SOME}, \texttt{<ALL}, etc. In the LP case (Fig.~\ref{fig:lp_architecture}) there are two such operators~--- a non-correlated and a correlated one.

Now, let us turn to internals. The implementation of LP operator in PosDB follows the block-based Volcano model. The core operator is decomposed into three main components~--- Modulator, Inverter, and Merger~--- with an auxiliary Proxy operator forwarding blocks between them. These operators exchange blocks of positions, synchronize outer and inner flows. Together, they coordinate caching, exclusion, and merging into the final output.

The data flow of the LP operator is illustrated in Fig.~\ref{fig:lp_implementation} and follows the pull-based model, where each operator requests blocks from its children on demand.
The Modulator slices blocks from the outer operator and forwards them through the left Proxy to the left non-correlated subplan, while caching them for later use.
The Inverter pulls blocks from the left subplan until enough input is accumulated. It then requests the cached blocks from the Modulator and performs exclusion to prepare the relevant data for the right correlated subplan.
The Merger pulls data both from the Inverter and the right correlated subplan, which receives the excluded outer blocks via the right Proxy. Finally, the Merger combines non-correlated results from the Inverter with the correlated results from the right subplan and returns the operator output in positional form.


Thus, we have proposed a novel operator for handling subqueries inside a position-enabled column-store. This operator accepts and returns data in positional form, extending the set of positional operators in the system. The strength of such systems lies in the ability to compose long chains of positional operators that propagate positions to the higher levels of a query plan. The longer the chains, the later the materialization occurs, and the greater the potential benefits of this approach.

\subsection{Cost Models}~\label{sec:model}

In order to assess the benefits of the proposed technique and to compare it with the naive approach, we have devised three simple cost models. These models will form the basis of a cost‑based optimizer that can determine whether the proposed approach is beneficial for a given query.

First, it is necessary to introduce several variables  for estimation:
\begin{itemize}
    \item $COSTP_{NC}$~--- cost of evaluating non-correlated predicate;
    \item $COSTP_{C}$~--- cost of evaluating correlated predicate;
    \item $NCFILTER$~--- the selectivity of the non-correlated predicate;
    \item $N$ and $M$~--- number of records in outer and inner tables, respectively.
\end{itemize}

For three cases described above the costs are the following:
\begin{enumerate}
    \item \textbf{Naive approach}: $N * M * (COSTP_{NC} + COSTP_{C})$ 
    \item \textbf{Rewritten query} (with caching): $N * COSTP_{NC} +  N * M * COSTP_{C}$
    \item \textbf{Proposed approach}: $N * COSTP_{NC} + (N * NCFILTER) * M * COSTP_{C}$
\end{enumerate}

These cost models imply that the performance of our approach depends on the selectivity of the non-correlated predicate. In the next section we are going to study its influence on the performance.

\section{Evaluation}

Experimental evaluation has been performed on a PC with the following characteristics: Archlinux 6.15.9-arch1-1, AMD Ryzen 5900X CPU, G Skill F4-3600C17-16GTZR RAM, Samsung SSD 990 PRO 2TB, gcc 15.1.1, PostgreSQL 17.5. We used TPC-H, with Scale Factor $0.2$ as the source data to run our experiments.

To evaluate our technique, we posed the following research questions:
\begin{enumerate}
    \item[RQ1] How is the performance of the proposed technique affected by non-correlated predicate selectivity?
    \item[RQ2] How does the performance of the proposed technique compare with industrial systems?
\end{enumerate}

\begin{lstlisting}[label={listing:eval-sql-orig}, caption={Evaluation query in the original form}]
SELECT P.partkey
    FROM part AS P
    WHERE P.retailprice < SOME (
        SELECT 3 * L.extendedprice * L.discount * L.tax
            FROM lineitem AS L
            WHERE L.suppkey = X or P.size = L.quantity
    );
\end{lstlisting}

\begin{lstlisting}[label={listing:eval-sql-rewritten}, caption={Evaluation query rewritten}]
SELECT P.partkey
    FROM part AS P
    WHERE P.retailprice < SOME (
        SELECT 3 * L.extendedprice * L.discount * L.tax
            FROM lineitem AS L
            WHERE L.suppkey = X
    ) OR P.retailprice < SOME (
        SELECT 3 * L.extendedprice * L.discount * L.tax
            FROM lineitem AS L
            WHERE P.size = L.quantity
    );
\end{lstlisting}

Our preliminary experiments confirmed that the hash-join based \texttt{IN} operator optimization employed by PostgreSQL shows a level of performance comparable to that of the proposed optimization (class 2). Therefore, we have decided to concentrate the evaluation efforts on queries belonging to the class 4. The query that was used for evaluation is presented in Fig.~\ref{listing:eval-sql-orig}, its rewritten form can be found in Fig.~\ref{listing:eval-sql-rewritten}. The evaluation plan includes exploring performance gains in scenarios of varying selectivity of the non-correlated subquery. For ease of such variation, parameter X has been introduced into the query. Values of X equal to 39, 1198, 734, 210, 451 deliver non-correlated subquery selectivity of 20\%, 40\%, 60\%, 80\% and 99.9\% respectively. The correlated subquery selectivity is 40\%. The evaluation measures the execution times of queries for each of the selectivity values in 5 different settings:

\begin{enumerate}
    \item Original query in PostgreSQL
    \item Rewritten query in PostgreSQL
    \item Original query in PosDB
    \item Rewritten query in PosDB with caching of NC subquery without LP
    \item Rewritten query in PosDB with caching of NC subquery and LP
\end{enumerate}

\begin{figure}[htp]
    \begin{center}
        \frame{\includegraphics[width=0.7\linewidth]{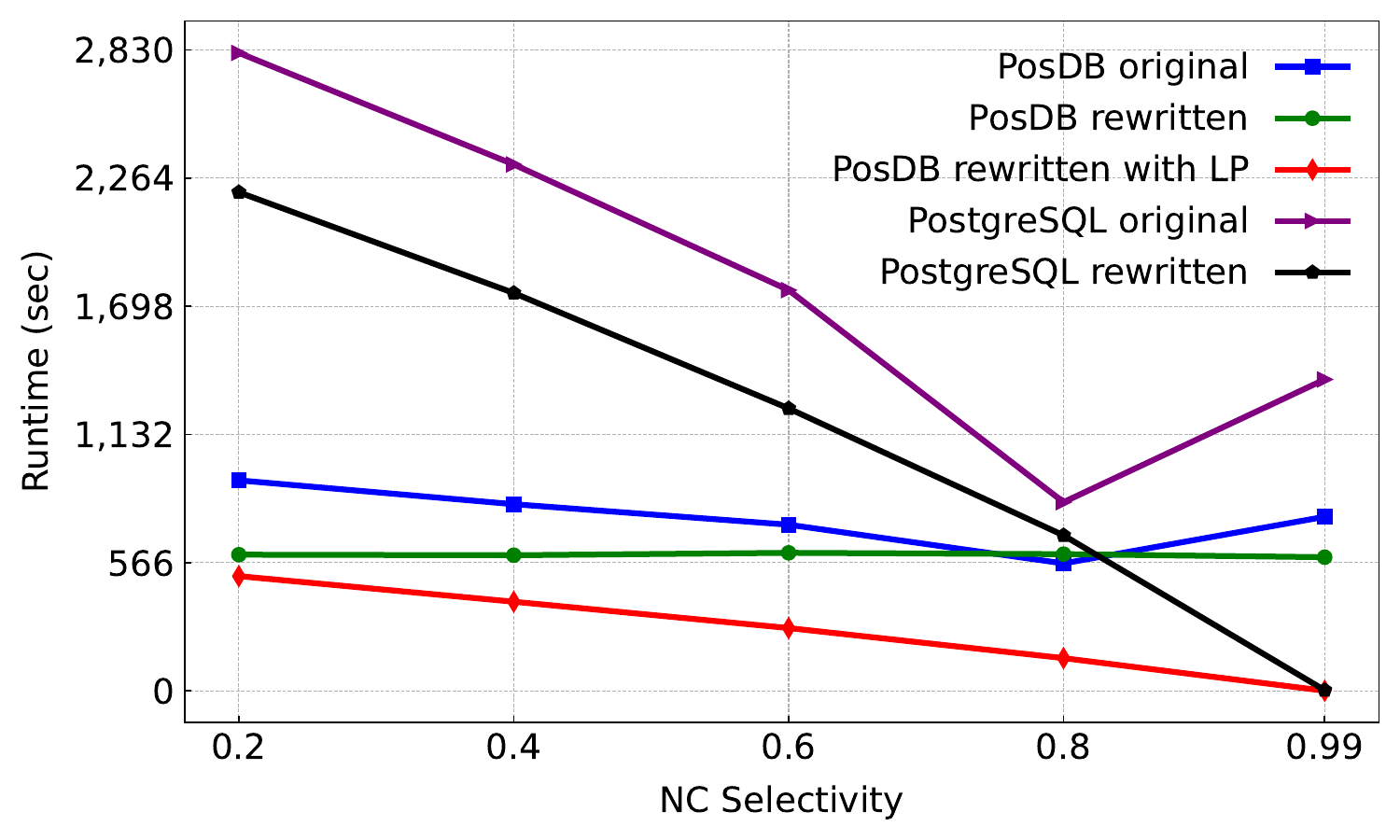}}
    \end{center}
    \caption{Execution times over NC selectivity}
    \label{plot:rq1}
\end{figure}

The results can be found in Fig.~\ref{plot:rq1}. For the rewritten query, it's apparent that for both PosDB with LP and PostgreSQL the execution times decrease linearly with increasing NC selectivity, converging to essentially 0 when selectivity is close to 100\%. This behavior aligns closely with the cost model discussed in Section~\ref{sec:model}, which predicts a direct relationship between performance and the proportion of tuples filtered by the non-correlated predicate. For PosDB without LP, it's clear that the execution time is more or less constant across all values of NC selectivity. This behavior is attributable to PosDB executing a hard‑coded query plan with no runtime optimization. Again, it is fully consistent with the cost models described in previous sections.

\begin{figure}[htp]
    \begin{center}
        \frame{\includegraphics[width=0.7\linewidth]{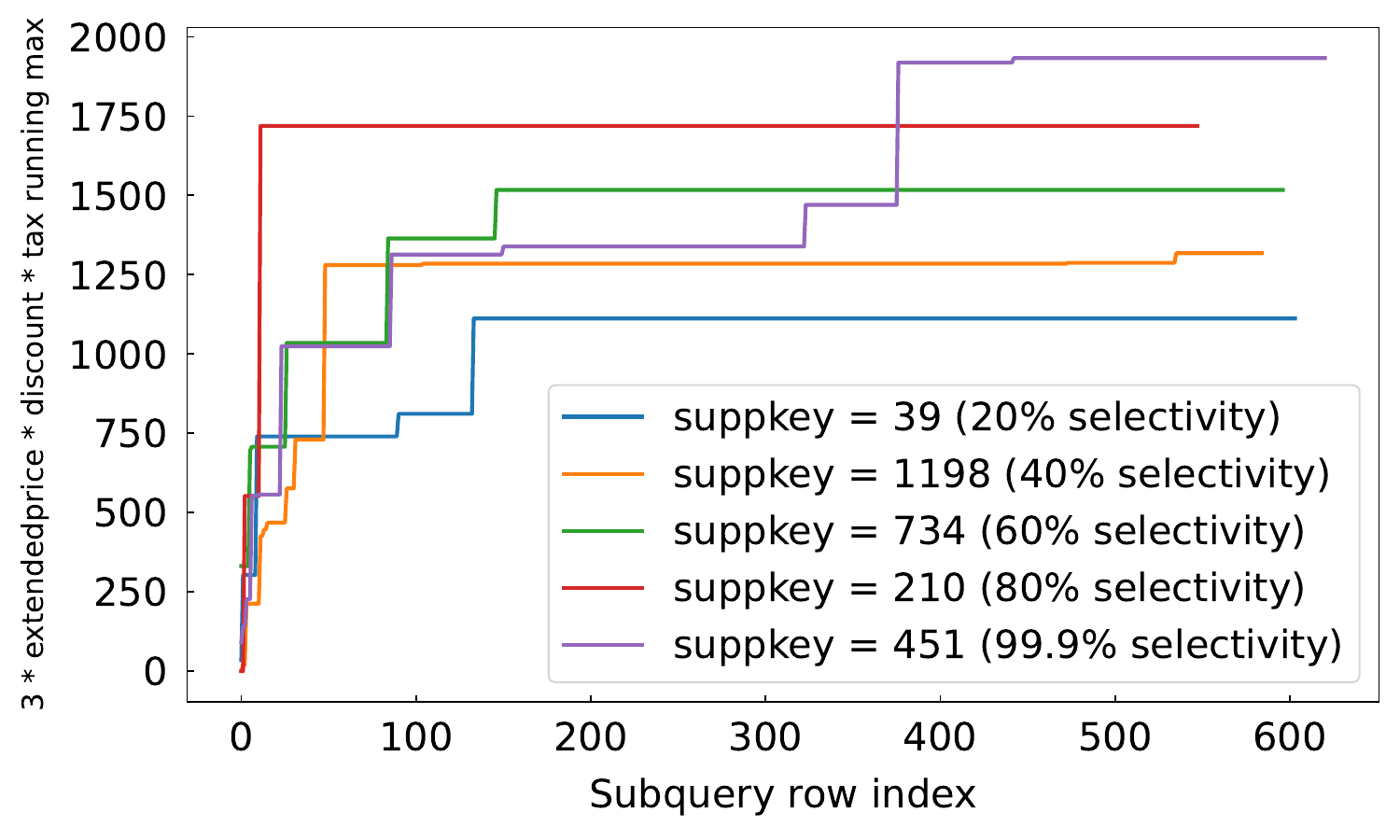}}
    \end{center}
    \caption{Running maximum of the expression value through subquery scan}
    \label{plot:supp}
\end{figure}

For the original query, several things are noteworthy. It's apparent that PostgreSQL does not implement the proposed rewriting of the query, since its performance here is different from its performance on the rewritten query. That is especially prominent in the 99.9\% NC selectivity case. That case is also very interesting because, despite the advantageous selectivity, it clearly has much higher runtime compared to that of 80\% NC selectivity, and comparable runtime to the query with 60\% NC selectivity. That happens because both PosDB and PostgreSQL implement early-exit optimization of SOME clause evaluation. For each row to the left, the processing stops immediately once a matching row on the right is found, and the scan of the right subquery is discontinued. Its execution time is greatly affected by how much of the query to the right has to be traversed before a matching row is found. Fig.~\ref{plot:supp} shows the running maximum of the calculated expression value in the order of NC subquery traversal for all the NC subqueries that were measured. Note that column ``retailprice'' of PART table has a uniform distribution in the 900--1900 range. It can be seen that, despite its eventual high selectivity value dictated by a high maximum value on the entire sequence, it only reaches that high value halfway through the scan. Which means that for a considerable part of the matching rows the execution would not be allowed to perform an early exit until it has evaluated and scanned most of the subquery. In a sense, the 99.9\% selectivity NC subquery can be considered ``unlucky'' for the early-exit optimization. On the other hand, the proposed approach doesn't have that vulnerability.

There's one more optimization approach that is very helpful in the query being considered that hasn't been evaluated here. It aims to speed up the evaluation of SOME predicate with a correlated subquery. It extends the idea of caching the maximum value of the evaluated expression that is used for non-correlated subqueries. Each time a correlated subquery has to be evaluated, the inputs to that query need to be calculated. The calculation takes the row from the outer query as input and produces some tuple of values. The result of the correlated subquery evaluation is then a deterministic function of that tuple. That allows the engine to cache the maximum value of the resulting subquery evaluation in a hashmap, using the tuple as the key. That allows the engine to bypass subquery evaluation whenever the outer row yields the same input tuple as one of the cached ones.

That technique is especially useful when the space of input tuples is small. In the query used above, both PART.SIZE and LINEITEM.QUANTITY are numeric values with a uniform distribution in the 1--50 range, which would only require the cache to store 50 items. In PosDB enabling that technique for the query above brings the execution time to values very close to 0. However, experiments demonstrated that PostgreSQL lacks this optimization.

\section{Conclusion}

In this paper we propose a novel technique for processing correlated SQL subqueries. Unlike many existing approaches, it is a simple method that does not require a significant redesign of a query engine. Next, we adapt the block‑based Volcano query processing model to support the technique and implement it inside a position‑enabled column-store as a position‑producing operator, thereby enabling late materialization when processing subqueries. The proposed approach can be employed by any column-store that allows flexible control over positions. Furthermore, our results are applicable to row stores as well.

To assess our technique, we evaluated it using PosDB and PostgreSQL. First, we studied the impact of non‑correlated predicate selectivity on the performance of our technique. This analysis, together with the simple cost models we also propose in this paper, forms the foundation of an optimizer that can decide when to apply our technique for a given query. Second, we compared our technique with the query plans produced by PostgreSQL. From these experiments we present an extensive discussion of the applicability boundaries of our technique.

\begin{credits}
\subsubsection{\ackname} 
We would like to thank Vladislav Makeev for his help with the preparation of the paper.
\end{credits}

\bibliographystyle{splncs04}
\bibliography{my-short}

\end{document}